# REVISTA ESPACIOS

# Sulfur emission reduction in cargo ship manufacturers and shipping companies based on MARPOL Annex VI

## Reducción de emisiones de azufre en fabricantes de buques y navieras según el Anexo VI del Convenio MARPOL

LONDOÑO-PINEDA, Abraham A.[1]
CANO, Jose A.[2]
PULGARIN-QUIROZ, Lissett [3]

**Abstract**
This article explores the challenges for the adoption of scrubbers and low sulfur fuels on ship manufacturers and shipping companies. Results show that ship manufacturers, must finance their working capital and operating costs, which implies an increase in the prices of the ships employing these new technologies. On the other hand, shipping companies must adopt the most appropriate technology according to the areas where ships navigate, the scale economies of trade routes, and the cost-benefit analysis of ship modernization.
**Keywords:** scrubbers; low sulfur fuels; ship manufacturers; shipping companies

**Resumen**
Este artículo explora los retos de implementar depuradores y combustibles con bajo contenido de azufre en fabricantes de barcos y navieras. Los resultados muestran que los fabricantes de barcos deben financiar su capital de trabajo y sus costos operativos, implicando un aumento de precios en los barcos que utilicen estas tecnologías. Las navieras deben elegir la tecnología más apropiada según las áreas donde navegan los barcos, economías de escala de las rutas comerciales y análisis de costo-beneficio de dichas tecnologías.
**Palabras clave:** depuradores; combustibles con bajo contenido de azufre; fabricantes de barcos; navieras

## 1. Introduction

Maritime transport is the one generating fewer CO2 emissions per tonne-mile compared to other means of transport like rail, ground, and air transportation (Lister et al., 2015). However, since the 1960s, the negative environmental effects generated by operational and accidental factors in maritime transportation have increased along with the rapid growth and annual movement of goods (Grant & Elliott, 2018; Ochoa, 2015). Some of these negative effects are related to sulfur oxides (SOx) and nitrogen oxides (NOx) emissions, air pollution, seas pollution, deaths of aquatic species and plankton due to oil and chemicals spills, garbage, and sediments

[1] Faculty of Economics and Administrative Sciences. Universidad de Medellin. alondono@udem.edu.co
[2] Faculty of Economics and Administrative Sciences. Universidad de Medellin. jacano@udem.edu.co
[3] Faculty of Economics and Administrative Sciences. Universidad de Medellin. lissettpulgarin@gmail.com

(House, 2004) Koski et al., (2017); Sato, (2018); Londoño, (2014); Londoño & Baena, (2017; Winther et al., (2014); Woodyard, ( 2004); de Oliveira et al.,( 2008).

This situation led to the adoption of the International Convention for the Prevention of Pollution from Ships (MARPOL) in 1973. This is the main convention related to the protection of the environment related to maritime transport and since then it has been continuously modified according to the realities of international trade and the impact on the environment (Lister et al., (2015); Walker et al., (2019). For this reason, Annex VI of MARPOL limited the maximum sulfur content of marine fuels in the Sulfur Emission Control Areas (SECA) to 0.1% from 2015 and 0.5% in 2020 for those areas outside of the SECA (Antturi et al., 2016); Nikopoulou, (2017); Olaniyi et al., (2018); van Hassel et al., (2016). SECA comprise the North Sea, Baltic Sea, the Caribbean Sea around Puerto Rico, Virgin Islands, and some nautical miles from Canada and the United States.

This circumstance demands an environmental commitment and investment in technologies to reduce sulfur emissions from ships based on exhaust gas cleaning systems ("scrubbers") and/or the incorporation of low sulfur fuels (Endres et al., 2018(); Gu & Wallace, (2O17); Han, (2010); Scott, (2017); Zheng & Chen, (2018) which leads to a series of financial effects for ship manufacturers and shipping companies renting such ships (Allal et al., 2019)  Solakivi et al., (2019).

## 2. Methodology

For the literature review, the databases of Scopus, Wos, and Science Direct were consulted using the search equation ("Shipowners" or "Shipping Companies") and ("Scrubbers" or "Low Sulfur Fuel"). This search provided 43 documents that address the sulfur emission reduction methods in the maritime sector, which consist of 34 journal articles, 5 conference proceedings, and 4 book chapters. Figure 1 shows the number of documents published per year, identifying that 79% of the documents were published between 2015 and 2019. This indicates that technologies and solutions for shipping companies, ship owners, and ship manufacturers that allow compliance with those proposed in chapter VI of MARPOL represent a current issue in the literature.

**Figure 1**
Documents per year

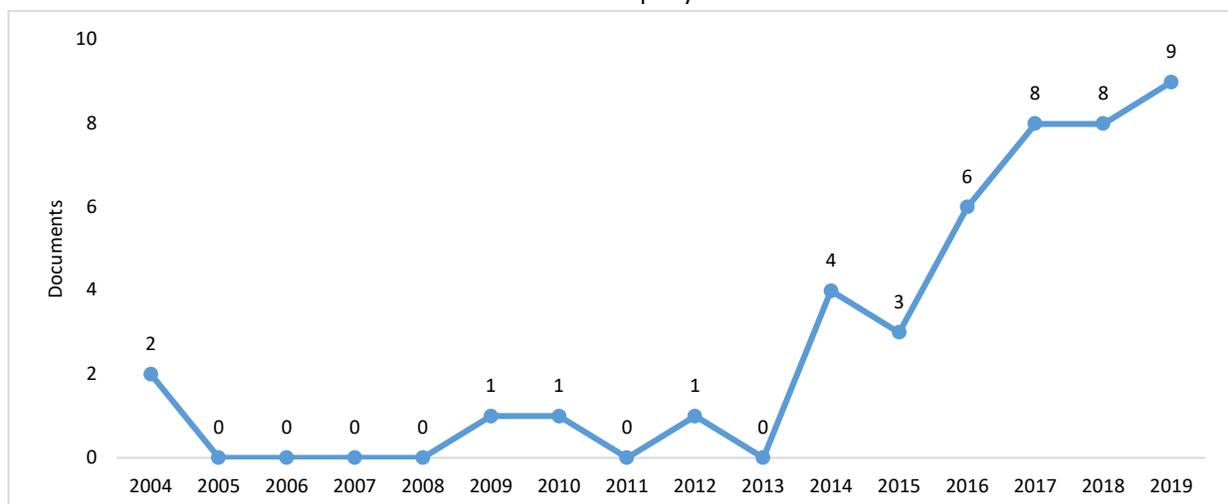

It is also identified that the journal Transportation Research Part D: Transport and Environment concentrates the largest number of documents published on the subject of this study, publishing 18% of the analyzed documents (8 documents). Authors with the largest number of publications on this topic are Thierry Vanelslander, Christa Sys, Kevin Cullinane, and Edwin van Hassel.

## 3. Results

The studies of Stevens et al. (2015), Armellini et al. (2018), Zhu et al. (2017), and Solakivi et al. (2019) argue that the use of scrubbers and diesel particulate filters is the best way to reduce sulfur emissions in ships, especially for shipping companies receiving significant operations flow through the SECA, since low sulfur fuels are usually more expensive (Gu & Wallace, 2017); Han, (2010); van Biert, et al., (2016); Zheng & Chen, (2018). On the other hand, the use of low sulfur fuels is more suitable for shipping companies operating with few ports calls in the SECA or operating outside the SECA (Acciaro, 2014); Calderón et al., (2016); Halff et al., (2019); Rehmatulla et al., (2017; Svanberg et al., (2018); Theocharis, et al., (2019); Vierth et al., (2015). Other factors like annual fuel consumption, fuel prices, and environmental regulation also influence deciding the best alternative to reduce sulfur emissions (Rehn et al., 2016).

Likewise, ship manufacturers face major challenges associated with high capital costs and the lack of comprehensive laws on air pollution, as it requires the commitment of actors like ports, governments, financial markets, and ship buyers, sellers and renters (Animah et al., 2018(; Cullinane & Bergqvist, (2014); Schinas & (2012). In addition, the need to improve port infrastructure is highlighted, especially for the application of technologies based on low sulfur fuels, as there are very few ports worldwide having an adequate infrastructure to provide this service (Cullinane & Cullinane, 2019) ; Holmgren et al., (2014); Kim & Seo, (2019); Notteboom & Vernimmen, (2009); Sanabria et al., (2015); Sheng et al., (2019).

Regarding the economic viability of scrubbers and low sulfur fuels, several analyses of long-term cash flows show that the costs of both types of technologies outweigh the benefits under current maritime regulation (Antturi et al., 2016(; Panasiuk and Turkina, (2015), which means that Annex VI of MARPOL runs the risk of non-compliance. As a consequence, the literature highlights the need to provide solutions that transcend the coercive nature (represented in fines, penalties), since this can affect both transport costs and the dynamism of international trade; alternatively, agreements and commitments between the different members of the maritime trade chain are required to jointly reduce sulfur emissions and contribute to the sustainability of maritime transport (Adland et al., 2017); Lähteenmäki-Uutela et al., (2017); Zis, (2019). Accordingly, the benefits of reducing sulfur emissions can be reflected in the improvement of the ecological footprint of the goods shipped. This also will increase the competitiveness and sustainability of supply chains that use ships equipped with scrubbers and/or powered by low sulfur fuels (Cano et al., 2015a; Cano et al., (2015b).

## 4. Conclusions

The main concern of ship manufacturers is related to sanctions that may be imposed if they do not adopt technologies to reduce sulfur emissions, obligating them to make large investments to comply with the normative requirements, impliying an increase in the sale price of vessels equipped with new technologies. In shipping companies is highlighted the use of opportunity cost analysis related to buy or rent ships with scrubbers or low sulfur fuels. These decisions depend mainly on factors such as ship navigation areas, scale economies of trade routes, and cost-benefit analysis of modernizing ships. Future research may propose different ways to improve the relationship and agreements between ship owners and ship renters in order to reduce sulfur emissions profitably and economically.

## Bibliographic references